# All-or-none switching of transcriptional activity on single DNA molecules caused by a discrete conformational transition

Ayako Yamada, Koji Kubo, Tonau Nakai and Kenichi Yoshikawa[*]

*Department of Physics, Graduate School of Science, Kyoto University & CREST, Kyoto 606-8503, Japan*

Kanta Tsumoto

*Department of Chemistry for Materials, Faculty of Engineering, Mie University, Tsu 514-8507, Japan*

Recently, it has been confirmed that long duplex DNA molecules with sizes larger than several tens of kilo-base pairs (kbp), exhibit a discrete conformational transition from an elongated coil state to a compact globule state upon the addition of various kinds of chemical species that usually induce DNA condensation. In this study, we performed a single-molecule observation on a

---

[*] Corresponding author. Fax: 81-75-753-3779.

E-mail address: yoshikaw@scphys.kyoto-u.ac.jp (K. Yoshikawa).

large DNA, Lambda ZAP II DNA (ca. 41 kbp), in a solution containing RNA polymerase and substrates along with spermine, a tetravalent cation, at different concentrations, by use of fluorescence staining of both DNA and RNA. We found that transcription, or RNA production, is completely inhibited in the compact state, but is actively performed in the unfolded coil state. Such an all-or-none effect on transcriptional activity induced by the discrete conformational transition of single DNA molecules is discussed in relation to the mechanism of the regulation of large-scale genetic activity.

An important goal in biological science is to clarify the mechanism of the self-regulation of transcriptional activity. Currently, many regulatory factors have been discovered, indicating that individual factors interact with certain regions by the recognition of specific sequences along long genomic DNA. It is considered that life is maintained through the use of very complicated networks composed of many key-lock relationships.[1,2] However, it is still unclear how, in living cells, it is possible to achieve a delicate control of numerous genes with such complicated networks. In a different field of research, cell biologists have been actively studying the relationship between morphology and function of chromatin, both in prokaryote[3] and eukaryote[4], and have shown that a marked morphological change is accompanied by a significant functional change such as differentiation, cell-cycle, and malignancy. Such large structural changes in chromatin have often been discussed in relation to the epigenetic modification of DNA and binding proteins like histone, in terms of methylation, acetylation, etc.[5,6] On the other hand, for the past decade it has been established that giant DNA molecules larger than several tens of kbp exhibit a large discrete conformational transition between elongated coil and folded compact states[7] upon the addition of various DNA condensing agents including biochemical species.[8,9] Thus, it is strongly suspected that this unique property of the folding transition of giant DNA should somehow be closely related to its biological function/activity. However, it have been some apparently conflicting studies on the biological activity of condensed DNA,[10-17] suggesting that the manner of "condensation" of DNA has not yet been adequately clarified biophysically. For example, it has been reported that "condensed DNA" shows rapid cyclization[10] and also higher transcriptional activity,[11] compared to the uncondensed DNA. In contrast, the inhibition of transcription[12] and depression of the activity of endonuclease[13] have also been reported. In the present article, we report our results in the direct observation of the on/off change of transcriptional activity induced by the folding/unfolding transition of a giant DNA in the presence of a biological polycation, spermine (4+).

We used Lambda ZAP II DNA (see Fig.1 right, bottom; Stratagene, CA), which includes a single T7 promoter region, as a template for transcription by T7 RNA polymerase (T7 RNAP; Invitrogen, MD). The transcriptional reactions were performed under conditions similar to those described in ref. 12, 100 μL of well-mixed reaction solution containing 5 mM dithiothreitol (DTT; provided with the polymerase kit), 0.5 mM each of ATP, CTP and GTP, 10 μM UTP (Roche Diagnostics, Switzerland), 0.5 μM BODIPY-UTP (Molecular Probes, Inc., OR), 0.3 μM DAPI (Wako Pure Chemicals Industries, Ltd., Japan) and 200 or 700 μM spermine tetrachloride (Nacalai Tesque, Japan) in T7 buffer (polymerase kit). After 80 or 10 ng/mL Lambda ZAP II DNA and 50 U of T7 RNAP were added under gentle stirring, the solution was incubated at 37ºC for 30 min. To avoid aggregation of DNA molecules, the concentration of DNA was adjusted to be very low; around 10 ng/mL. Although RNAP activity on a single DNA molecule has been observed in a variety of adroit ways in previous trailblazing researches,[18-21] here we used very simple and concise procedure to visualize both of the DNA conformation and its transcripts: after the transcriptional reaction, we elongated DNA molecules on the glass surface by gentle shear stress.[22] Both the Lambda ZAP II DNA and RNA molecules were visualized with a Carl Zeiss fluorescence microscope (Axiovert 200 TV) equipped with a 100× oil-immersed lens and a high-sensitivity Hamamatsu SIT TV camera. DAPI bound to DNA and BODIPY incorporated into RNA by polymerization was excited at wavelengths of 365 and 535 nm, respectively. We also used an atomic force microscope (AFM) equipped with a fluorescence microscope (AFM; NBV100, FM; IX70, Olympus Optical Co. Ltd., Japan) to observe the same DNA and RNA molecules. AFM was performed in Tapping Mode™ in aqueous solution.

Figure 1(A) shows typical fluorescence images of Lambda ZAP II DNA; (a) in an unfolded elongated coil state and (b) in a folded compact state. In the lower pictures in Fig. 1(A), the intensity distribution of fluorescence emitted from DAPI is given as a quasi-3D representation. In Fig. 1(A), the unfolded stretched state of DNA (a) is obtained as described above. The red spot in the upper left picture in Fig. 1(A) indicates the active production of RNA, where fluorescence from BODIPY is observed. In contrast, no red fluorescence is emitted from the folded compact DNA, as shown in the upper right picture in Fig. 1(A).

Figure 1(B) shows experimental data on the dependence of transcriptional activity on the spermine concentration through conventional test-tube measurement, *i.e.*, the activity is measured as the ensemble average of many DNA molecules. The experimental condition in this measurement is essentially the same as in the single-DNA observation. The results in Fig. 1(B) indicate that the change in transcriptional activity around [spermine (4+)] = 400 μM is rather steep, although it is not clear, based solely on the observation of the ensemble of DNAs, whether the change in transcriptional activity is on/off type or cooperative and continuous type. The single-chain observation clarifies that this activity is completely inhibited for all of the compact DNA molecules.

To measure the transcription along a DNA with higher resolution, we performed AFM/FM observation. Figure 2 shows the AFM image of an unfolded DNA together with the fluorescence microscopic image of RNA labeled with BODIPY. Due to the resolution limit with optical observation, the fluorescence observation of BODIPY only gives a blurred image around the region that is white in the AFM image. Based on a comparison of the AFM and fluorescence images, in Fig. 2(c) we show a schematic representation of our observation. Although we could not recognize individual RNAP/RNA molecules, the white part in AFM is attributable to RNAP molecules that stay near the promoter in transcription.

Table I summarizes the results regarding transcriptional activity with the single-chain observation. At a spermine concentration of 700 μM, the activity of transcription was completely inhibited and all of the DNA molecules assumed a folded compact state. In contrast, at 200 μM, where DNA molecules show an unfolded coil conformation, most of the DNAs exhibit the transcriptional activity. It is now clear that individual DNA shows on/off switching in transcriptional activity that is accompanied by a folding transition.

The above result indicates that the transcriptional activity of an individual chain is strongly related to its conformation, *i.e.* to a drastic change in its density. The density ratio between the globule and coil states $\rho_{globule}/\rho_{coil}$ equals the inverse ratio of the cube of the respective radii $R^3_{coil}/R^3_{globule}$. The coil radius is given as $R_{coil} \sim \lambda N^{3/5}$ under good solvent conditions for a chain,[23] where $\lambda$ is the persistence length (about 150 bp for DNA at room temperature) and $N$ is defined as the number of independent segments of length $\lambda$ per contour length $L$ (40820 bp for Lambda ZAP II DNA). On the other hand, the globule size $R_{globule}$ is proportional to $(\lambda s N)^{1/3}$ where $s$ is the cross-sectional area of the chain (3 nm$^2 \cong$ 27.5 bp$^2$). The density ratio $\rho_{globule}/\rho_{coil}$ can thus be estimated as $s^{-1}\lambda^2 N^{4/5}$. For the giant DNA adopted in this experiment, this ratio is ~10$^5$. It has been confirmed from the measurement of the Brownian motion of individual giant DNA molecule that the relative ratio of the hydrodynamic radii is on the order of 10$^4$-10$^5$ on giant DNA.[24] The high density present in a globular conformation should prevent the binding of enzymes (here RNA polymerase) to DNA, which could lead to a total inhibition of transcriptional activity. Table I indicates that the relative activity decreased from 0.60 at 200 μM spermine to 0.00 at 700 μM spermine. The abrupt change in density, with a discrete conformational transition, is the direct cause of the on/off switching of transcriptional activity. The present study shows that DNA transcription undergoes an on/off change that is accompanied by the conformational transition of DNA, which is induced by a small change in the solution; *i.e.*, the spermine concentration in the present study. It has been reported that a small change in concentration such as of a 1:1 salt, RNA and ATP, can cause a large discrete transition in the conformation of giant DNA.[24] Such a conformational transition in giant DNA can cover a large genetic region, possibly including tens or hundreds of genes. Thus, it is strongly suspected that such a conformational transition corresponds to the opening/closing of a folder that contains several genes.

In the present study, we have shown that transcription is completely inhibited upon compaction. In contrast to this inhibitory effect, it has been reported that transcriptional activity was promoted by the condensation on small circular plasmid DNA molecules.[11] Circular DNA with superhelicity tends to undergo a continuous-type folding transition[25] in contrast to the discrete conformational transition of linear giant DNA[7]: circular DNA exhibits a stepwise collapse, which causes local stress along double-stranded DNA chains.

It is expected that the generation of such local stress in circular DNAs induced by the collapsing transition may be associated with the enhancement of enzymatic activity, through weakening of the double-strand structure. A similar discussion may be applicable to the genetic activity in a loop attached to a scaffold in giant genetic DNA in chromatin,[4] although the other factors such as topoisomerases and SMC proteins, etc., are also considerable. Since the folding transition is a characteristic event generated on a large scale along a giant DNA, it may play a role in the large-scale regulation of multiple genes[24], complementing key-lock interactions in individual genes.

The authors are grateful to Dr. Naoko Makita for providing helpful advice on the experimental methods, Mr. Hiroyuki Kitahata for help on analyzing video images and Dr. Tatsuo Akitaya and Dr. Damien Baigl for their kind comments. This work was supported by a Grant-in-Aid for the 21 century COE, "Center for Diversity and Universality in Physics".

TABLE I.

Transcriptional activity at different spermine concentrations obtained from single-DNA measurement

| [Spermine] | Number of active DNA | Total number of DNA* | Fraction of active DNA |
|---|---|---|---|
| 200 μM | 47 | 79 | 0.60 |
| 700 μM | 0 | 103 | 0.00 |

* Unfragmented DNA only.

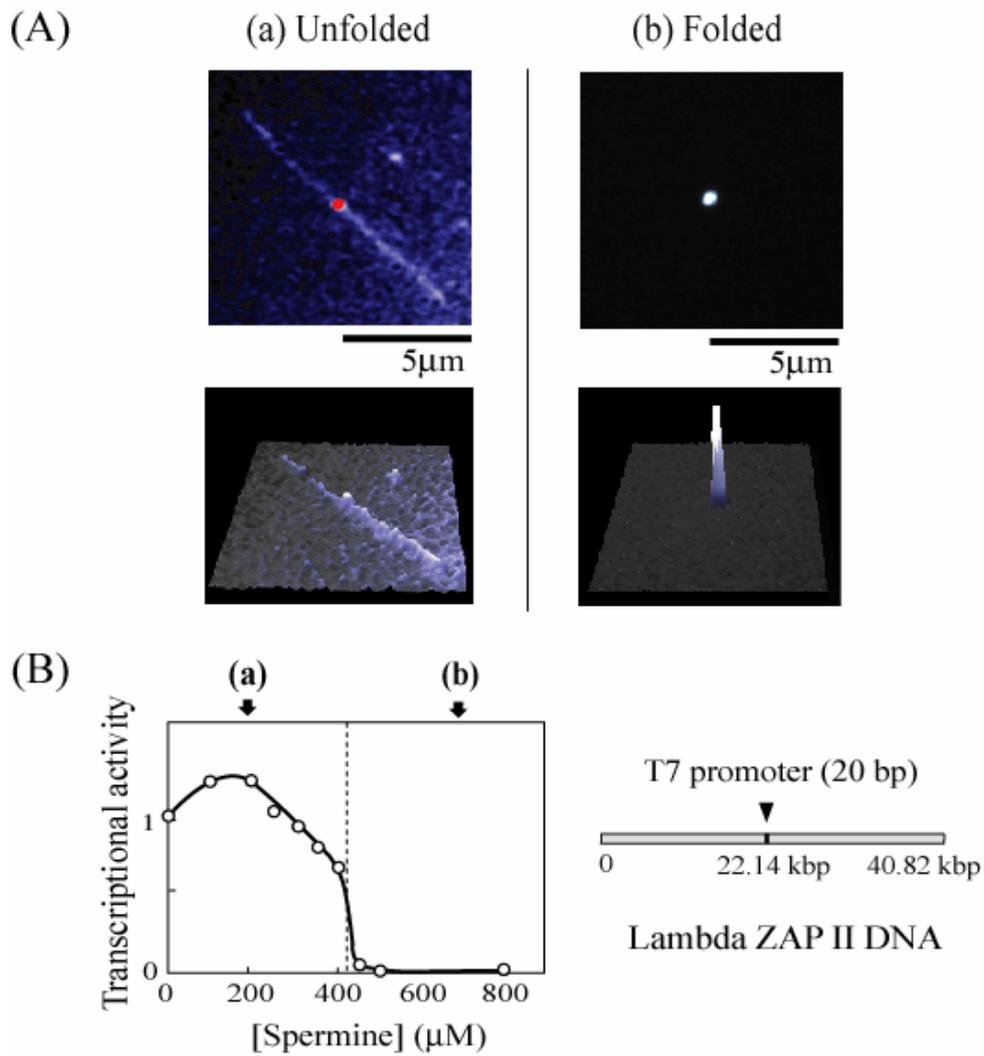

FIG. 1: (A) Fluorescence microscopic images of single Lambda ZAP II DNA at (a) [spermine (4+)] = 200 μM and (b) [spermine (4+)] = 700 μM. Upper pictures: blue and red colors correspond to the fluorescence from labeled DNA and RNA (transcripts), respectively. Lower pictures: quasi-three-dimensional representations of the spatial distribution of fluorescence intensity on DNA. (B) Transcriptional activity vs. concentration of spermine (4+). The vertical broken line indicates the region of the conformational transition as evaluated from the fluorescence microscopic observation of bulk DNA solution. (Modified from the original figure in ref. 12)

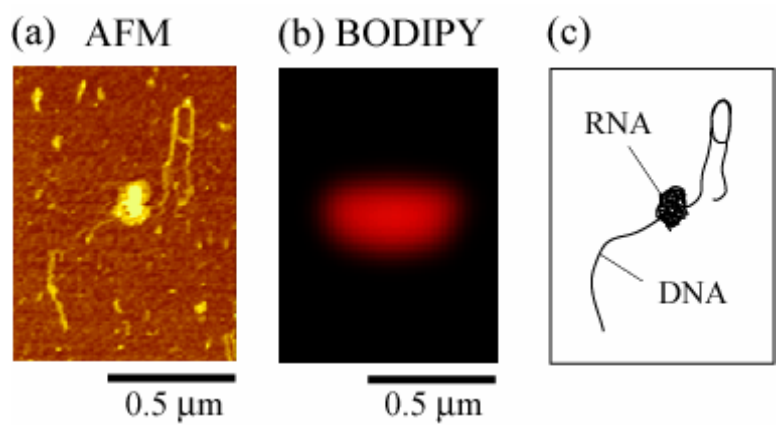

FIG. 2: A single DNA molecule transcribing RNA, as observed simultaneously by (a) AFM and (b) FM. (c) Schematic representation.